\documentclass[aps,prl,reprint,superscriptaddress,nofootinbib,longbibliography]{revtex4-2}

\usepackage{amsmath,amssymb,bm}
\usepackage{graphicx}
\usepackage[hidelinks]{hyperref}
\hypersetup{hidelinks}
\usepackage{xcolor}
\usepackage{booktabs}
\usepackage{array,tabularx,microtype}
\usepackage{placeins}
\usepackage{siunitx}
\usepackage{mathtools}
\usepackage[normalem]{ulem}

\newcommand{\e}{\mathrm{e}}
\newcommand{\chipn}{\chi_{\rm PN}}

\begin{document}

\title{Learning Inspiral–Merger–Ringdown Waveforms from a Post-Newtonian Baseline}

\author{Arghya Chattopadhyay}
\affiliation{Physics Department, University of Puerto Rico Mayag\"uez, Puerto Rico 00681, USA}
\email{arghya.chattopadhyay@upr.edu}

\author{Shilpa Kastha}
\affiliation{Saha Institute of Nuclear Physics, 1/AF Bidhannagar, Kolkata 700064, India}
\affiliation{Homi Bhabha National Institute, Training School Complex, Anushaktinagar, Mumbai 400094, India}
\email{shilpa.kastha@saha.ac.in}

\date{\today}

\begin{abstract}
Modeling the full inspiral–merger–ringdown signal requires combining analytically controlled inspiral physics with the nonlinear strong-field information supplied by numerical relativity. We represent the numerical relativity contribution beyond an analytic inspiral waveform as residual amplitude and phase corrections. Retaining the leading-order frequency-domain amplitude and the $3.5$ Post-Newtonian TaylorF2 phase, we use a Kolmogorov–Arnold network to learn these residual corrections from SXS waveforms. After training, the learned corrections are stored as explicit spline functions, so waveform evaluation no longer requires the network itself. On $75$ simulations excluded from training and model selection, the model achieves a median flat-noise mismatch of $2.7\times10^{-5}$. Our results demonstrate a machine learning driven waveform modeling strategy in which numerical relativity augments, rather than replaces, analytically known waveform structure.
\end{abstract}

\maketitle

{\it Introduction}: Gravitational wave (GW) inference from binary black hole (BBH) coalescences requires waveform models that accurately describe the inspiral, merger, and ringdown (IMR) while remaining efficient for repeated likelihood evaluations. The inspiral admits a controlled perturbative description~\cite{Blanchet2024}, whereas the late-inspiral and merger dynamics must be informed by numerical relativity (NR)~\cite{LehnerPretorius2014}. Modeling the full signal therefore requires connecting  
the analytically tractable inspiral to the nonlinear strong-field regime.

Existing waveform models address this problem in complementary ways. Effective-one-body models recast the binary dynamics into a resummed effective description~\cite{Buonanno:1998gg,Buonanno:2000ef,PhysRevD.108.124035}, with selected ingredients informed by NR~\cite{Pan:2011gk}; phenomenological models construct analytic amplitude and phase representations calibrated to hybrid and NR waveforms~\cite{Ajith:2009bn,Husa:2015iqa,Khan:2015jqa, PhysRevD.103.104056}; and surrogate models construct reduced representations of NR waveforms and interpolate them across the source-parameter space~\cite{Blackman:2015pia,Blackman:2017pcm,Varma:2019csw}. 
 
Machine-learning methods have also been applied to waveform modeling. Neural networks and Gaussian-process regression have been used to interpolate waveform families or accelerate existing model performances~\cite{Doctor:2017csx,Chua:2018yng,Setyawati:2019xzw,Schmidt:2020kxc,Khan:2020fso,Thomas:2022qss}, while more data-driven approaches construct surrogates from NR or NR-surrogate waveforms~\cite{Khan:2021mnb,Freitas:2024nrsurnn}.
Closest in spirit to the present work, physics-informed approaches have used neural networks to learn corrections to binary dynamics~\cite{Keith:2021} and to Post-Newtonian (PN) waveforms using NR information~\cite{Yoo:2025yhl}. 
These developments motivate a complementary question. Can we retain the analytically controlled inspiral and learn only the missing strong-field correction from NR? This factorization preserves the analytically known physics baseline while isolating the information supplied by NR.

Here we extend this residual-learning strategy to the IMR waveform of aligned-spin binaries. We use a Kolmogorov-Arnold network (KAN)~\cite{liu2025kan} to learn the amplitude and phase corrections required to reproduce NR while retaining the analytically controlled inspiral structure explicitly. Because the trainable functions are represented by splines, the learned correction can be reduced to a finite collection of spline-product terms and evaluated independently of the KAN training framework. The final model thus provides an explicit numerical description of the NR information missing from the analytic waveform baseline.

We apply this construction to the dominant $(2,2)$ mode of quasi-circular, non-precessing BBHs. After factoring out the overall total mass and luminosity distance scalings, we take the Newtonian-order ($0$ PN) stationary-phase amplitude and the aligned-spin TaylorF2 phase through $3.5$PN order~\cite{Buonanno2009,Bohe2013,Mishra:2016whh} as the analytic baseline.
The KAN learns only the logarithmic-amplitude and intrinsic phase residuals from SXS NR simulations spanning \(1\leq q\leq8\) and \(|\chi_{1,2}|\leq0.8\). 
On 75 independent configurations excluded from training and model selection, the resulting waveform reaches a median time- and phase-maximized flat-noise mismatch of $2.7\times10^{-5}$. This demonstrates that a compact learned residual can extend a simple inspiral model through merger and ringdown with high fidelity.

An important feature of this construction is that the learned correction remains an explicit analytic function after training. Rather than encoding the waveform dependence implicitly in a dense multilayer perceptron, the KAN organizes the residual into a small number of spline functions of frequency and intrinsic parameters whose contributions can be inspected, differentiated, and tested individually. This provides a compact intermediate description between a fixed phenomenological ansatz and a fully black-box neural surrogate, while retaining the flexibility needed to represent the NR strong-field corrections. The same framework can also be adapted with little structural change to alternative analytic backbones or different NR waveform families, making it a reusable strategy for constructing new physics-informed waveform models.

{\it Physics-Informed waveform construction:  }We consider quasi-circular, non-precessing BBHs with component masses $m_1\ge m_2$, and dimensionless spins $\chi_1$ and $\chi_2$ aligned or anti-aligned with the orbital angular momentum. We define the total mass $M=m_1+m_2,$ mass ratio $q = m_1/m_2$, the symmetric mass ratio $\eta=\frac{m_1m_2}{M^2},$ and the luminosity distance $D_L$ and collect the intrinsic parameters as $\lambda=(\eta,\chi_1,\chi_2)$. Introducing $M_{\rm sec}=\frac{GM}{c^3}, D_{\rm sec}=\frac{D_L}{c},$ and $x=M_{\rm sec}f,$ the frequency-domain waveform is written as
\begin{equation}
\widetilde h(f;\lambda,M,D_L)=
\frac{M_{\rm sec}^{2}}{D_{\rm sec}}\,
 a(x;\lambda)\,\e^{i\phi(x;\lambda)},
\label{eq:scaling}
\end{equation}
The total mass and luminosity distance scalings are thus factored out explicitly, leaving the model to describe the dimensionless amplitude and phase.

Rather than learning the full waveform, we retain the analytically known inspiral structure and model only the residual corrections required to reproduce NR. We take the leading-order frequency-domain amplitude $a_N(x,\eta)$~\cite{Cutler:1994ys} and the aligned-spin TaylorF2 phase $\Psi_{\rm TF2}(x,\boldsymbol{\lambda})$, through $3.5$PN order~(Appendix B, of \cite{PhysRevD.93.044007}), as the analytic baseline. This factorization was guided by our experience during training, where we found that learning became more effective when the analytic baseline already captured the known inspiral structure. The logarithmic-amplitude residual is defined as $
d_{\rm NR}(x,\boldsymbol{\lambda})
=
\ln\left[
|\widetilde h_{\rm NR}(x,\boldsymbol{\lambda})|/
a_N(x,\eta)
\right].$ Here the corresponding frequency domain NR waveform is $\tilde h_{\rm NR}(x,\lambda)$.

For the phase, using the convention that the analytic inspiral waveform carries the factor $\exp[-i\Psi_{\rm TF2}]$, we define the raw phase-residual
$$
\Delta\phi_{\rm NR}(x,\boldsymbol{\lambda})
=
\arg\!\left[
\widetilde h_{\rm NR}(x,\boldsymbol{\lambda})
\right]
+
\Psi_{\rm TF2}(x,\boldsymbol{\lambda}).
$$
For each NR waveform, this residual contains an arbitrary contribution from the choice of time and phase origin, which is affine in frequency. We therefore define the intrinsic phase residual by
\begin{align}
    r_{\rm NR}(x,\boldsymbol{\lambda}) = \Delta\phi_{\rm NR}(x,\boldsymbol{\lambda})-2\pi x{\tau}^{\rm align}_i - \varphi_i^{\rm align},
\end{align}
where $\tau_i^{\rm align}$ and $\varphi_i^{\rm align}$ are simulation-specific alignment coefficients obtained from a low-frequency alignment fit. These coefficients are used only to construct the intrinsic training target $r_{\rm NR}$; they are neither supplied to nor predicted by KAN.

KAN is therefore trained on $d_{\rm NR}$ and $r_{\rm NR}$. The learned corrections are combined with the analytic baseline to reconstruct
\begin{align}
\widetilde h_{\rm KAN}(f)
=&
\frac{M_{\rm sec}^{2}}{D_{\rm sec}}\,
a_N(x,\eta)\,
\exp\left[d_{\rm KAN}(x,\boldsymbol{\lambda})\right] \nonumber \\ 
& \times \exp\left[i\Phi_{\rm KAN}(x,\boldsymbol{\lambda})\right],
\end{align}
where the intrinsic phase is
$$
\Phi^{\rm int}_{\rm KAN}(x,\boldsymbol{\lambda})
=
-\Psi_{\rm TF2}(x,\boldsymbol{\lambda})
+
r_{\rm KAN}(x,\boldsymbol{\lambda}),
$$
and the full waveform phase is then
\begin{align}
\Phi_{\rm KAN}(x,\boldsymbol{\lambda})
=
\Phi^{\rm int}_{\rm KAN}(x,\boldsymbol{\lambda})+
2\pi x \tau_c
+
\phi_c.  
\end{align}
with $\tau_c\equiv t_c/M_{\rm sec}$. The signs of the last two terms follow the Fourier convention adopted here. The coalescence time \(t_c\) and phase \(\phi_c\) remain free waveform parameters, and are unrelated to $\tau_i^{\rm align}$ and $\varphi_i^{\rm align}$.

To organize the intrinsic-parameter dependence, we introduce, $\chi_s=(\chi_1 + \chi_2)/2, \chi_a=(\chi_1 - \chi_2)/2, \delta=\sqrt{1-4\eta}$ together with the spin combinations $\chipn=\left(1-\frac{76}{113}\eta\right)\chi_s+\delta\chi_a$~\cite{PhysRevD.88.064007}, and 
$\zeta=\delta\chi_a$. The combination $\chi_{\rm PN}$ captures the leading aligned-spin dependence of the PN phase, while $\zeta$ retains sensitivity to the antisymmetric spin degree of freedom. We further introduce $u=A_v[(\pi x)^{1/3}]$ and $
w=\frac{x}{x_{\rm RD}(\lambda)}$ where $A_v$ maps the PN velocity variable onto the spline domain.

We define the ringdown frequency scale as $
x_{\rm RD}\equiv Mf_{220}
=\frac{M}{M_f}\frac{M_f\omega^{R}_{220}}{2\pi},
$ where $M_f$ and $\omega^{R}_{220}$ are the remnant mass and the real frequency of the fundamental $(2,2,0)$ Kerr quasinormal mode. The remnant mass and spin are estimated using the radiated energy based on the nonspinning fit of Ref.~\cite{Husa:2015iqa} supplemented by a simplified spin correction. And $\omega^{R}_{220}$ is then obtained from Ref.~\cite{Berti:2005ys}. The coordinate $u$ follows the PN inspiral scaling, whereas $w$ aligns the merger-ringdown structure across the binary parameter space.

For either residual $F\in\{r_{\rm KAN},d_{\rm KAN}\}$, the trained representation is written as,
\begin{align}
F={}&f_0(u)+g_0(w)+\sum_{\rho=1}^{R_u} f_\rho(u)
S^{(u)}_\rho(\widehat\eta,\widehat\chi_{\rm pn})
 z^{(u)}_\rho(\widehat\zeta)\nonumber\\
&+\sum_{\rho=1}^{R_w} g_\rho(w)
S^{(w)}_\rho(\widehat\eta,\widehat\chi_{\rm pn})
 z^{(w)}_\rho(\widehat\zeta).
\label{eq:architecture}
\end{align}
We have further affinely rescale $\chi_{\rm PN}$, $\zeta$, and the PN velocity to their respective spline domains, denoting the rescaled intrinsic coordinates by $\widehat{\chi}_{\rm PN}$ and $\widehat{\zeta}$. We use $R_u=R_w=6$, which provides a good compromise between accuracy and compactness, yielding a model with $4192$ trainable coefficients. The one-dimensional functions $f_\rho$, $g_\rho$ and $z_\rho$ are represented by cubic B-splines, while $S_\rho$ are tensor-product spline surfaces in the two-dimensional intrinsic-parameter subspace. The two sums separately encode frequency dependence organized by the inspiral-scaled coordinate $u$ and the ringdown-scaled coordinate $w$ with their variation across the intrinsic binary parameters carried by the spline factors. After training, both the residual corrections are represented by a finite sum of spline products. The two representations together contain $24$ spline-product terms, whose coefficients fully specify the trained correction.

{\it Numerical-relativity data and calibration: } The model is constructed from version $3.0.0$ of the SXS catalog~\cite{SXSCatalogPaper_3, SXSCatalogData_3.0.0}. We select non-precessing BBH simulations with $1\leq q\leq 8$, $|\chi_{1,2}|\leq 0.8$ and the reported eccentricity below $10^{-3}$. One waveform is excluded because its post-merger duration is insufficient for the conditioning procedure described below, leaving $460$ systems. From the dominant SXS strain mode we construct the face-on dimensionless waveform
\begin{equation}
h(\tau)=\sqrt{\frac{5}{4\pi}}\frac{r h_{22}(\tau)}{M},
\qquad
\tau=\frac{t}{M},
\end{equation}
which is subsequently transformed to the frequency domain to obtain the amplitude and phase residuals used for training.

Because the SXS waveforms are finite time series, their frequency-domain representation requires conditioning that suppresses boundary artifacts without modifying the strong-field portion of the signal. After removing the early-time portion affected by junk radiation and uniformly resampling the waveform, we apply a smooth Planck window~\cite{McKechan:2010kp} whose late-time turn-off follows the measured post-merger amplitude decay. The window is exactly unity over
\begin{align}
\tau\in
[\tau_{\rm peak}-50,\,
 \tau_{\rm peak}+4\tau_{\rm RD}],
\end{align}
where $\tau_{\rm RD} = 1/(M |{\rm Im}~\omega_{220}|)$ is the damping time of the remnant's fundamental $(2,2,0)$ quasinormal mode.

The frequency interval used for fitting is chosen independently of the time-domain window. For each binary, $x_{\rm start}$ is defined as the median instantaneous frequency of the retained $h_{22}$ mode over its first $200M$. We set $x_{\rm low}=2x_{\rm start}$ to suppress finite-start effects and $x_{\rm high}=1.25x_{\rm RD}$ to exclude the resolution-sensitive high-frequency tail. We verify the robustness of these choices, together with the other conditioning parameters, below.

Varying the initial cut, sampling interval, zero padding, late-time cutoff, and frequency boundaries over a representative validation subset changes the affine-projected phase by at most $1.18\times10^{-3}\,\mathrm{rad}$ in root-mean-square over the common analysis interval. For comparison, the $90$th-percentile highest-versus-next-highest SXS resolution difference is $1.90\times10^{-3}\,\mathrm{rad}$ in a $50$-system resolution study. The conditioning and boundary sensitivity is therefore smaller than the characteristic finite-resolution variation of the NR waveforms.

To distinguish performance in well-sampled regions of parameter space from that in regions with sparse NR coverage, the $460$ systems are partitioned geometrically in the intrinsic space $(\eta, \chi_{PN}, \zeta)$ before training. Near-duplicate configurations are grouped together, and a space-filling sample of $280$ systems defines the training set. The remaining well-supported configurations are divided into $75$ validation and $75$ held-out test systems, while $30$ configurations with comparatively sparse local NR coverage form a separate stress set. The validation set is used for model selection; the held-out test and stress sets enter neither coefficient fitting nor reduction of the final representation.

During training, we form the phase error $e_i(x)=r_{\rm KAN}(x,\lambda_i)-r_{\rm NR}^{(i)}(x)$ for each training waveform $(i^{th})$. At every loss evaluation and for each phase-loss weighting, we independently remove from $e_i$ its weighted least-squares projection onto the affine basis $\{1,2\pi x\}$, using that component’s frequency mask and weights. The projected error enters the corresponding phase loss, making the optimization invariant under arbitrary relative time and phase shifts. For optimization, the frequency weights are smoothly turned on using a $C^2$ roll-on from $Mf=0.0059$ to 
$Mf=0.0070$, avoiding a sharp low-frequency boundary in the loss. This training-time projection is distinct from the unweighted low-frequency alignment used to construct $r_{\rm NR}^{(i)}$. The stored coefficients $\tau_i^{\rm align}$ and $\phi_i^{\rm align}$ do not enter the training tensors and thus are not predicted by the KAN. See the Spplemental Material for details of the NR pre-processing, parameter-space partition, KAN training and model reduction.

\begin{figure}[t]
\centering
\includegraphics[width=0.48\textwidth]{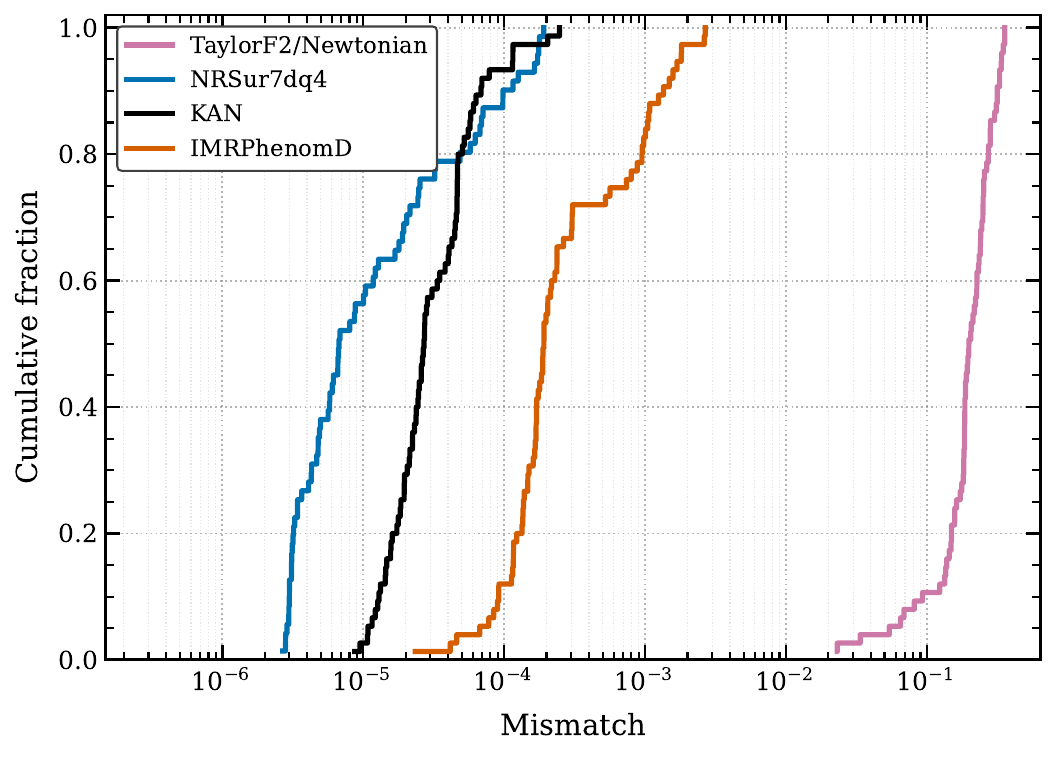}
\caption{Empirical cumulative distributions of the time- and phase-maximized flat-noise mismatch with respect to SXS waveforms. KAN, IMRPhenomD, and the TaylorF2/Newtonian-amplitude backbone are evaluated on the $75$ held-out test configurations; NRSur7dq4 is shown for the $71$ test configurations admitting a comparison within its strict calibration domain. For each system, the mismatch is computed on a uniform frequency grid over the largest continuous intersection of the trustworthy frequency supports of the waveforms being compared, with flat frequency weighting. The SXS and NRSur7dq4 waveforms are conditioned in the time domain and Fourier transformed once, whereas KAN, IMRPhenomD, and the analytic backbone are evaluated directly in the frequency domain.}
\label{fig:CDF_all}
\end{figure}

\begin{figure*}[t]
\centering
\includegraphics[width=0.45\textwidth]{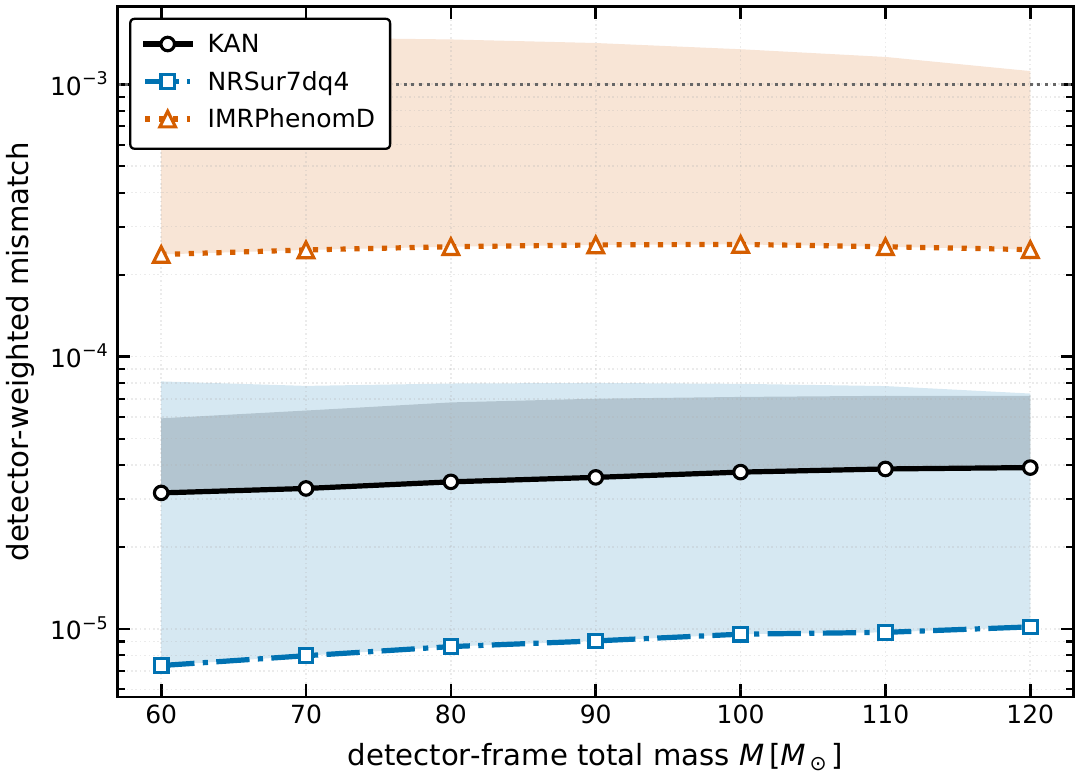}
\hfill
\includegraphics[width=0.45\textwidth]{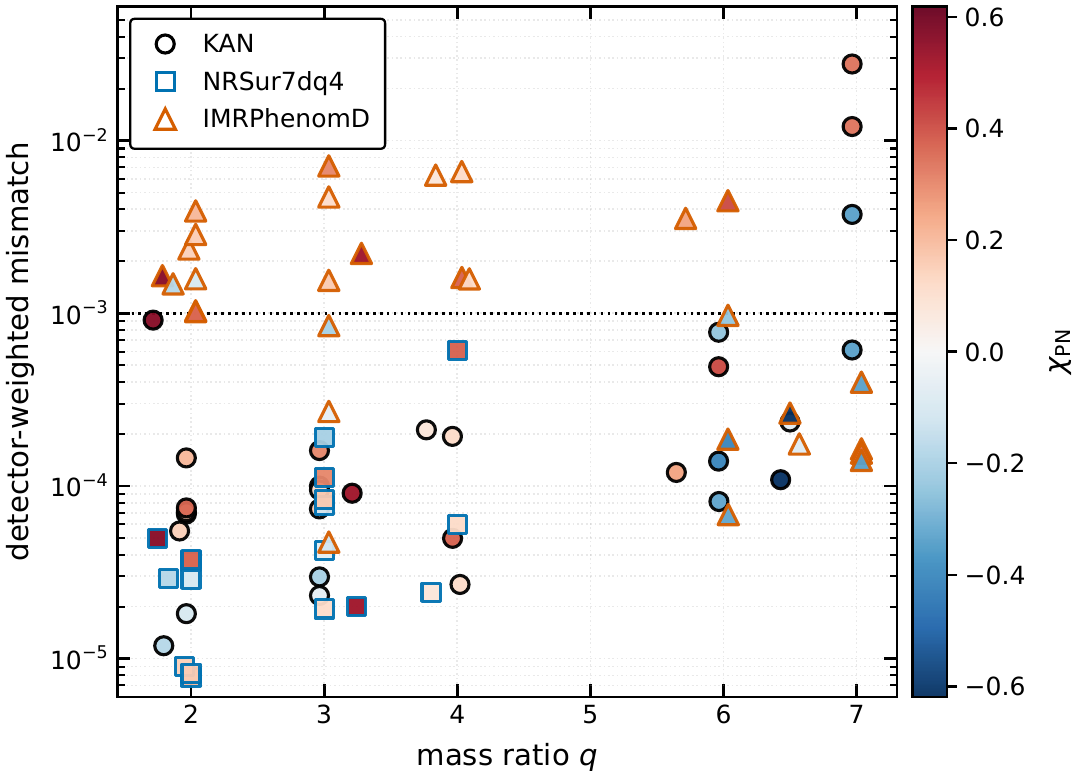}
\caption{Detector-weighted, time- and phase-maximized mismatches with respect to SXS waveforms, computed using the Advanced LIGO design power spectral density with $f_{\rm low}=20\,\mathrm{Hz}$. Left: mismatch as a function of detector-frame total mass for the $71$ held-out test configurations within the strict NRSur7dq4 calibration domain. Curves show the median mismatch for KAN, NRSur7dq4, and IMRPhenomD, and the shaded regions extend from the median to the corresponding $90$th percentile over the 71 systems. Right: mismatches for the 30 configurations in the sparse-coverage stress set at $M=60\,M_\odot$, shown as a function of mass ratio $q$, with color bar indicating $\chi_{\rm PN}$. NRSur7dq4 results are shown only for the 18 configurations of the $30$ within its strict calibration domain.
}
\label{fig:detector_weighted}
\end{figure*}

{\it Waveform accuracy: }We quantify the agreement between an SXS waveform $(h_{\rm SXS})$ and a model waveform $(h_{\rm model})$ using the time- and phase-maximized match
\begin{equation}
\mathcal{M}
=
\max_{t_c,\phi_c}
\frac{
\langle h_{\rm SXS}\vert h_{\rm model}\rangle
}{
\sqrt{
\langle h_{\rm SXS}\vert h_{\rm SXS}\rangle
\langle h_{\rm model}\vert h_{\rm model}\rangle
}
}\label{eq:mismatch},
\end{equation}
with ${\rm mismatch}=1-\mathcal{M}$,
where
\begin{equation}
\langle h_1\vert h_2\rangle
=
4\,{\rm Re}
\int_{f_{\rm low}}^{f_{\rm high}}
\frac{
\widetilde h_1(f)\widetilde h_2^{*}(f)
}{
S_n(f)
}\,{\rm d}f .
\label{eq:inner_product}
\end{equation}
The relative phase is maximized analytically and the time shift numerically. Our primary benchmark adopts, $S_n(f)=1$, providing a detector-independent measure of waveform agreement. For the quoted flat-noise mismatches, we impose a reference lower-frequency cutoff $Mf\geq0.0059$, corresponding to $20\,\mathrm{Hz}$ for a $60\,M_\odot$ binary. For each binary, all models entering a given comparison are evaluated against the same SXS waveform over the intersection of their valid frequency intervals, using identical frequency samples and weights.

FIG.~\ref{fig:CDF_all} shows the mismatch distributions for the 75 held-out test configurations, with NRSur7dq4 evaluated on the 71 systems within its strict calibration domain. For each configuration, the TaylorF2/Newtonian baseline, KAN model, NRSur7dq4~\cite{NRSur7dq4}, and IMRPhenomD~\cite{Husa:2015iqa, Khan:2015jqa} are compared against the same SXS waveform over their common trustworthy frequency interval. The TaylorF2/Newtonian baseline has a median mismatch of $\sim 0.2$ demonstrating the importance of the learned strong-field correction. The KAN reduces the median to \(2.7\times10^{-5}\), with a 90th percentile of $6.64\times10^{-5}$. IMRPhenomD yields a median mismatch of \(1.91\times10^{-4}\) and a $90$th percentile of \(1.31\times10^{-3}\). In a system-by-system comparison, the KAN has lower mismatch than IMRPhenomD for $73$ of the $75$ held-out systems, corresponding to $97.3\%$ of the test set. NRSur7dq4 gives the lowest median mismatch, $6.75\times10^{-6}$, and is more accurate than the KAN for \(70.4\%\) of the systems.

We next assess detector-weighted accuracy using the Advanced-LIGO design power spectral density~\cite{Shoemaker2009ALIGOSensitivity} with $f_{\rm low}=20\,{\rm Hz}$ (left panel, FIG.~\ref{fig:detector_weighted}). For detector-frame total masses of 60, 90, and 120,$M_\odot$, the KAN median mismatches are $3.2\times10^{-5}$, $3.6\times10^{-5}$, and $3.9\times10^{-5}$, respectively. NRSur7dq4 remains at the $\sim10^{-6}$ level, while IMRPhenomD remains at a few $10^{-4}$. The relative performance therefore remains unchanged under detector weighting. For the separately defined $30$-system stress sample, at $M=60\,M_\odot$, the KAN has a median mismatch of $1.04\times10^{-4}$, while the 90th percentile rises to $1.2\times10^{-3}$ (right panel of FIG.~\ref{fig:detector_weighted}). The larger mismatches quantifies the loss of interpolation accuracy in sparsely sampled regions of the training domain and provides an empirical diagnostic of the model's domain of validity.

{\it Discussion: }We have constructed an inspiral–merger–ringdown waveform model using KAN framework, in which the analytically controlled inspiral structure is retained explicitly, while NR supplies only the residual amplitude and phase corrections. After training, these corrections are represented by an explicit low-rank spline-product model that can be evaluated independently of the KAN framework. This demonstrates that the NR contribution can be isolated as a correction to the analytic inspiral baseline, providing a physics-factored construction of the complete waveform.

The comparisons with existing waveform models clarify the role of this construction. Within the NRSur7dq4 calibration region, the dedicated NR surrogate remains closer to SXS for most systems, as expected. The present KAN instead preserves an explicit analytic baseline, extends over the aligned-spin range considered here to $q\leq8$, and improves on IMRPhenomD for all the test configurations. Its purpose is therefore not to replace an NR surrogate, but to showcase that the NR information missing from an analytic waveform can itself be isolated and compressed into an explicit correction.

The stress sample exposes the principal limitation, with interpolation accuracy degrading where the available NR coverage becomes sparse. The current model is restricted to the dominant $(2,2)$ mode of quasi-circular, non-precessing binaries. An advantage of the present formalism is that the learned correction remains explicit, compact, and modular, so that the analytic backbone and the NR training set can be modified without changing the overall construction. Extending the same physics-factored strategy to higher harmonics, precession, and eccentricity will therefore test whether this separation between analytically controlled physics and learned strong-field corrections remains effective for more general waveform families. More broadly, the results support a data-assisted modeling strategy in which NR supplements, rather than replaces, the physics that is already known analytically. {\it The waveform model, scripts used to generate the figures reported in this Letter are publicly available at Ref.~\cite{ChattopadhyayKasthaKAN2026}}.

\section{acknowledgments}

This work was initiated during a visit to the University of Mississippi in January 2026, when severe winter weather brought activities in Oxford to a standstill. We thank Anuradha Gupta and Breese Quinn for their generous hospitality during the visit. S.K. also acknowledges the hospitality of Prof. Sudhir Malik at the University of Puerto Rico at Mayag\"uez. The work of AC is partially supported by U.S.A. National Science Foundation Award OAC-$2334265$. AI tools, including Anthropic’s Claude and OpenAI’s ChatGPT, were used to assist with code
development and language refinement during the preparation of this work.

\bibliographystyle{apsrev4-2}
\bibliography{references}

\onecolumngrid
\graphicspath{{figures/}}
\renewcommand{\thefigure}{S\arabic{figure}}
\renewcommand{\thetable}{S\arabic{table}}
\renewcommand{\theequation}{S\arabic{equation}}
\newcommand{\dlnA}{\delta\!\ln A}
\newcommand{\rd}{\rm RD}



\section{Supplemental Material for\\
``Learning Inspiral--Merger--Ringdown Waveforms from a Post-Newtonian Baseline''}
This supplemental write-up is intended to shed light on the technical details of converting numerical-relativity (NR) waveforms into trustworthy frequency-domain residual targets. We start by explaining the data pre-processing done for training the machine learning model and with the explanation of how the constrained Kolmogorov-Arnold representation is constructed, how it is initialized and optimized, and how the trained representation is reduced to a deployable waveform model.

\section{Data processing and trustworthy numerical-relativity targets}
\label{sec:data}

\subsection{Catalogue selection and waveform normalization}

We use version 3.0.0 of the Simulating eXtreme Spacetimes (SXS) catalogue \cite{SXSCatalogPaper_3,SXSCatalogData_3.0.0}.  We select binary-black-hole simulations that are nonprecessing, have reported eccentricity below $10^{-3}$, and satisfy
\begin{equation}
  1\le q\equiv \frac{m_1}{m_2}\le 8,
  \qquad |\chi_1|,|\chi_2|\le 0.8,
\end{equation}
while consistently canonicalizing to $m_1\ge m_2$ with the spin labels exchanged together with the masses when necessary. This selection yields $461$ archived systems.  One system, $SXS:BBH:0202$, is removed because the numerical record terminates too soon after merger to accommodate the merger-protecting late-time taper described below. The final data set therefore contains $460$ waveforms.

The SXS data provide the spin-weighted-spherical-harmonic strain modes
in dimensionless NR units\footnote{The SXS waveforms are provided in
	geometrized units, $G=c=1$, with the initial total mass $M$ setting the
	characteristic length and time scale. Thus $t/M$ is dimensionless, and
	the asymptotic strain mode is conventionally supplied as
	$r h_{\ell m}/M$, which is likewise dimensionless. Physical masses and
	distances are restored only when constructing the detector-frame
	waveform.}. For a general observer direction, the complex strain may be decomposed as
\begin{equation}
	h(\tau;\iota,\varphi)
	=
	\sum_{\ell,m}
	\frac{r h_{\ell m}(\tau)}{M}\,
	{}_{-2}Y_{\ell m}(\iota,\varphi),
	\qquad
	\tau\equiv\frac{t}{M},
\end{equation}
where $\iota$ is the inclination between the line of sight and the
orbital angular momentum, and $\varphi$ specifies the azimuthal
orientation. In the present work we restrict attention to the dominant
$(2,2)$ mode and adopt a face-on orientation, $\iota=0$. For this
orientation,
\begin{equation}
	{}_{-2}Y_{22}(0,\varphi)
	=
	\sqrt{\frac{5}{4\pi}}\,e^{2i\varphi}.
\end{equation}
The arbitrary azimuthal origin may be chosen as $\varphi=0$, so that the
complex waveform used throughout this work becomes
\begin{equation}
	h(\tau)
	=
	\sqrt{\frac{5}{4\pi}}\,
	\frac{r h_{22}(\tau)}{M},
	\qquad
	\tau\equiv\frac{t}{M}.
	\label{eq:faceon}
\end{equation}
Note that the division by $M$ removes the remaining overall mass dimension. All preprocessing is then performed in the dimensionless variables $(\tau,x)$, while the physical mass and distance scalings are restored when constructing the detector-frame waveform as needed.

The analytic reference retained in the final model consists of the Newtonian stationary-phase amplitude \cite{Cutler:1994ys},
\begin{equation}
 a_N(x,\eta)=\sqrt{\frac{5}{24}}\,\pi^{-2/3}\sqrt{\eta}\,x^{-7/6},
 \label{eq:newtonianamp}
\end{equation}
and the aligned-spin TaylorF2 phase through 3.5PN order \cite{Buonanno2009,Bohe2013,Mishra:2016whh}. These analytic expressions define the baseline relative to which the SXS amplitude and phase corrections are constructed. The raw phase residual retains the arbitrary time and phase origins of each numerical simulation. A constant phase offset contributes a frequency-independent term, while a time translation gives a contribution linear in frequency. We therefore fit the affine form
\begin{equation}
	\Delta\phi_{\rm NR}^{(i)}(x)
	\simeq
	2\pi x\,\tau_i^{\rm align}
	+\phi_i^{\rm align}
\end{equation}
over a low-frequency inspiral interval by unweighted least squares, and define
\begin{equation}
	r_{\rm NR}^{(i)}(x)
	=
	\Delta\phi_{\rm NR}^{(i)}(x)
	-
	2\pi x\,\tau_i^{\rm align}
	-
	\phi_i^{\rm align}.
\end{equation}
The low-frequency interval provides a smooth reference for fixing these two gauge freedoms without allowing the merger-ringdown structure to influence the alignment. The coefficients $\tau_i^{\rm align}$ and $\phi_i^{\rm align}$ are used only to construct the training targets and are neither supplied to nor predicted by the KAN.

The NR targets learned by the KAN are therefore not the complete waveform but the corrections
\begin{align}
	d_{\rm NR}(x,\lambda)&=\ln\!\left[\frac{|\widetilde h_{\rm NR}(x,\lambda)|}{a_N(x,\eta)}\right],\\
	\Delta\phi_{\rm NR}(x,\lambda)&=\arg\widetilde h_{\rm NR}(x,\lambda)+\Psi_{\rm TF2}(x,\lambda),\\
	r_{\rm NR}(x,\lambda)&=\Delta\phi_{\rm NR}(x,\lambda)-2\pi x\tau_i^{\rm align}-\phi_i^{\rm align}.
	\label{eq:targets}
\end{align}

\subsection{Adaptive time window and the merger-protecting prescription}

The main challenge in Fourier transforming a finite NR waveform is that the numerical signal begins and ends at finite times. An abrupt truncation in the time domain produces spectral leakage in the frequency domain, while an overly aggressive taper can remove genuine late-inspiral, merger, or ringdown signal. We therefore condition each waveform only outside the portion that we regard as physically essential. After discarding the early-time region contaminated by junk radiation, the waveform is uniformly resampled with $\Delta\tau=0.5$ and multiplied by a smooth Planck window $W(\tau)$ constructed using the Planck-taper\footnote{A fixed turn-off can overlap the physical merger or early ringdown for simulations with a short post-merger record, whereas attaching an analytic continuation would introduce additional modeled information into a target that we want to obtain directly from NR. The adaptive amplitude-triggered Planck window was chosen to avoid both possibilities while leaving the merger and early ringdown unchanged.} functional form of \cite{McKechan:2010kp}. If $\tau_{\rm peak}$ denotes the time at which $|h_{22}|$ reaches its maximum, we impose
\begin{equation}
	W(\tau)=1\quad\text{for}\quad
	\tau\in[\tau_{\rm peak}-50,\,\tau_{\rm peak}+4\tau_{\rm RD}] .
	\label{eq:protected}
\end{equation}
This condition leaves unchanged the final $50M$ before the amplitude peak and the first four damping times of the fundamental ringdown mode. The late-time taper may begin only after this protected interval and once the waveform amplitude has fallen below $10^{-2}$ of its peak value. Whenever sufficient post-merger data are available, the
record is retained until the amplitude reaches the $10^{-4}$ level, with a minimum turn-off width of $10M$. At early times, a nominal $600M$ turn-on is shortened when necessary so that $W(\tau)=1$ by $\tau_{\rm peak}-50M$. For all 460 retained simulations the window is therefore exactly unity throughout the protected interval.

Both $\tau_{\rm RD}$ and the ringdown frequency scale used later in the
KAN are obtained from the fundamental $(2,2,0)$ quasinormal mode of the
remnant black hole. Writing the final-to-initial mass ratio as
$\mu_f=M_f/M$, we use \cite{Berti:2005ys}
\begin{equation}
	x_{\rm RD}\equiv Mf_{220}
	=\frac{M_f\omega^R_{220}}{2\pi\mu_f},
	\qquad
	\frac{\tau_{\rm RD}}{M}
	=\mu_f\frac{2Q_{220}}{M_f\omega^R_{220}} .
	\label{eq:qnmconversion}
\end{equation}
Here $M_f\omega^R_{220}$ and $Q_{220}$ are the dimensionless real
frequency and quality factor of the Kerr quasinormal mode. 

\begin{figure}[t]
	\centering
	\includegraphics[width=0.95\linewidth]{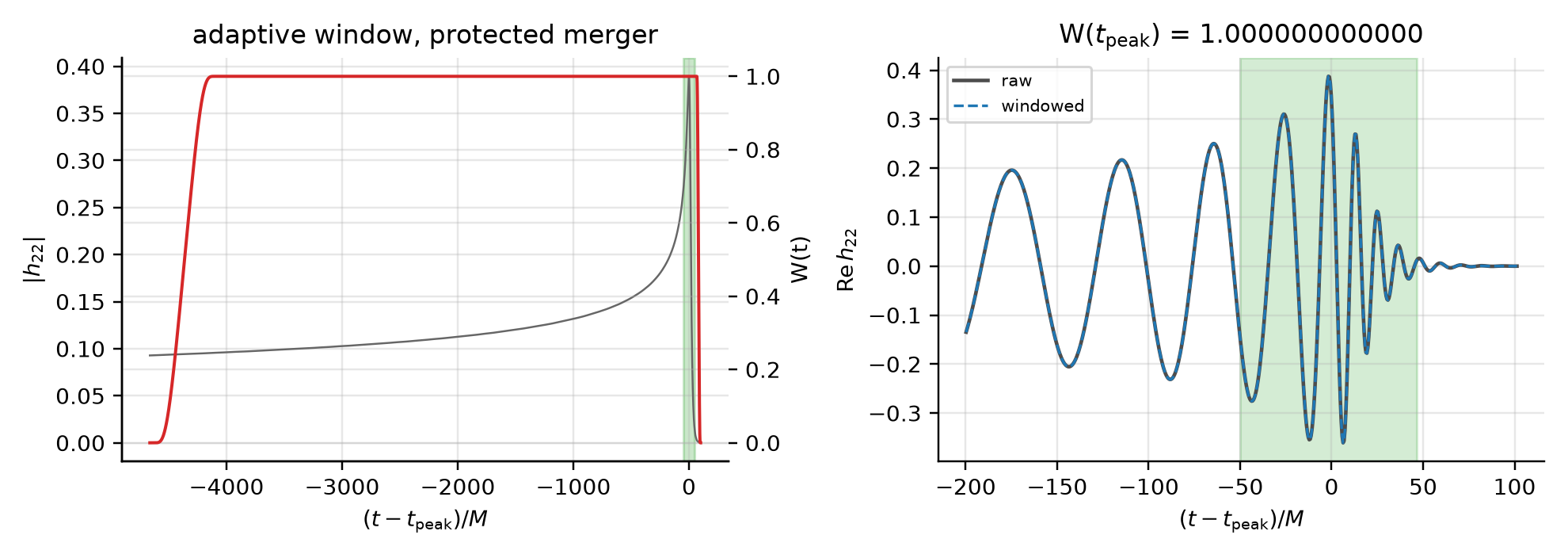}
	\caption{Representative adaptive time-domain conditioning.  The window remains exactly unity throughout the protected late-inspiral, merger, and early-ringdown interval of Eq.~\eqref{eq:protected}.  The late-time taper begins only after the protected interval and after the ringdown amplitude has sufficiently decayed.}
	\label{fig:window}
\end{figure}

\subsection{Fourier convention, trustworthy frequency support, and NR resolution}

For the uniformly sampled complex waveform, we construct the frequency-domain signal using the discrete fast Fourier transform (FFT), the numerical implementation of the discrete Fourier transform. With the NumPy convention,
\begin{equation}
	H(x)=\Delta\tau\,\mathrm{FFT}[W(\tau)h(\tau)],
\end{equation}
where the forward transform carries the phase convention $\exp(-2\pi i f t)$. With this convention the complex SXS $h_{22}$ mode appears on the negative-frequency branch. We therefore construct the positive-frequency waveform used by the model as
\begin{equation}
	\widetilde h_{\rm SXS}(x)
	=\frac{1}{2}H(-x)^*,
	\qquad x>0 .
	\label{eq:fftconv}
\end{equation}
The complex conjugation maps the physical $m=2$ contribution onto the positive-frequency convention adopted for the frequency-domain model, while the factor $1/2$ converts the two-sided complex representation to the corresponding one-sided waveform convention.

The trustworthy frequency interval is determined separately from the time-domain taper.  For each binary, we estimate the initial dimensionless gravitational-wave frequency $x_{\rm start}$ from the median instantaneous $(2,2)$ frequency over the first $200M$ of the retained waveform. The NR residual is then used for training only over
\begin{equation}
	x_{\rm low}=2x_{\rm start},
	\qquad
	x_{\rm high}=1.25x_{\rm RD}.
	\label{eq:validband}
\end{equation}
The lower boundary provides a buffer against Fourier contamination associated with the finite beginning of the numerical waveform. Note that this does not imply that the physical inspiral begins at $2x_{\rm start}$. Below this boundary, the TaylorF2/Newtonian waveform
continues to provide the analytic reference, while the numerical-relativity residual is used only where the finite SXS record supplies a stable frequency-domain target.

The frequency boundaries must also exclude regions where changes in numerical conditioning or NR resolution become comparable to the physical residual that the KAN is intended to learn. We tested the robustness of the adopted frequency interval in two ways.
First, we varied the main preprocessing choices, including the junk-radiation cut, sampling interval, zero padding, late-time cutoff, and nearby frequency boundaries. These changes modify the affine-projected phase by at most $1.18\times10^{-3}$ rad over the
common comparison band. Second, we compared the highest and next-highest available SXS resolutions for a stratified sample of 50 simulations to estimate the size of finite-resolution differences. We find
\begin{align}
	\mathrm{median}[1-\mathcal M]_{\rm res}&=5.02\times10^{-7}, &
	p_{90}[1-\mathcal M]_{\rm res}&=4.00\times10^{-6},\\
	\mathrm{median}[\mathrm{RMS}(\Delta\phi)]_{\rm res}
	&=3.47\times10^{-4}\ {\rm rad}, &
	p_{90}[\mathrm{RMS}(\Delta\phi)]_{\rm res}
	&=1.90\times10^{-3}\ {\rm rad},\\
	\mathrm{median}[\mathrm{RMS}(\Delta\ln A)]_{\rm res}
	&=3.92\times10^{-4}, &
	p_{90}[\mathrm{RMS}(\Delta\ln A)]_{\rm res}
	&=2.19\times10^{-3}.
	\label{eq:resstats}
\end{align}
These are finite-resolution differences rather than a universal NR
error floor. For the 18 systems with three usable resolutions, the
difference decreases with increasing resolution in 17 cases. The same
study motivates the choice $x_{\rm high}=1.25x_{\rm RD}$ as well. The
90th-percentile phase difference remains near $2\times10^{-3}$ rad up
to $1.25x_{\rm RD}$, but increases to $5.81\times10^{-3}$ rad at
$1.4x_{\rm RD}$ and $2.22\times10^{-2}$ rad at $1.5x_{\rm RD}$.

\subsection{Coverage-aware partition of parameter space}

To test genuine interpolation rather than memorization of closely related NR simulations, the 460 usable systems are first grouped into 251 near-duplicate clusters, with each cluster assigned wholly to a single subset. The partition is constructed using only the intrinsic coordinates
\begin{equation}
	\chipn=\left(1-\frac{76}{113}\eta\right)\chi_s+\delta\chi_a,
	\qquad
	\zeta=\delta\chi_a,
	\qquad
	\delta=\sqrt{1-4\eta},
	\label{eq:spincoords}
\end{equation}
where $\chipn$ captures the leading aligned-spin PN dependence
\cite{PhysRevD.88.064007}. A space-filling training set is selected in $(\eta,\chipn,\zeta)$, while the remaining configurations are separated according to their local coverage into interpolation-supported and sparse-coverage samples. This allows us to assess ordinary interpolation separately from performance in poorly sampled regions, without using waveform mismatch or residual information to define the split.

\begin{figure}[t]
\centering
\includegraphics[width=0.98\linewidth]{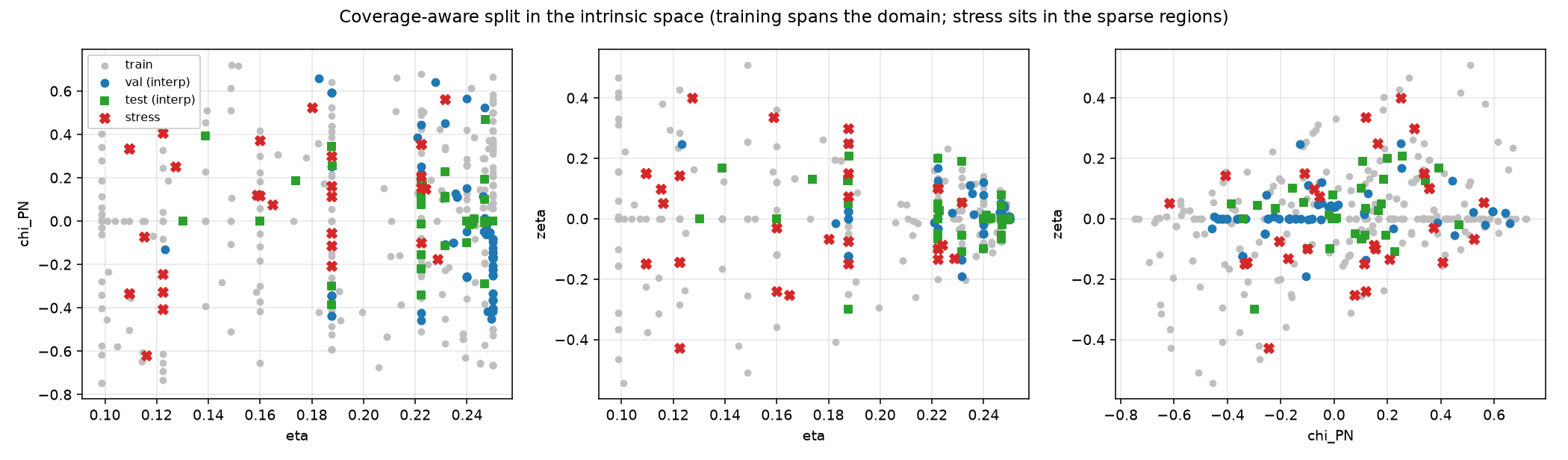}
\caption{Coverage-aware partition in $(\eta,\chipn,\zeta)$.  The 30-system stress sample remains inside the occupied catalogue domain but lies in regions with substantially poorer local training coverage.}
\label{fig:split}
\end{figure}

The resulting sets contain 280 training, 75 validation, 75 held-out interpolation-test, and 30 stress systems. The validation set is used for model selection, while neither the test nor stress set enters coefficient fitting or selection. Figure~\ref{fig:split} shows the resulting
parameter-space coverage.

\section{KAN and spline representation}
\label{sec:kan}

\subsection{From the Kolmogorov-Arnold theorem to a trainable representation}

The architectural motivation comes from the Kolmogorov-Arnold representation theorem.  In one common form, a continuous function of $n$ variables on a compact domain can be written as
\begin{equation}
 F(x_1,\ldots,x_n)=\sum_{q=1}^{2n+1}\Phi_q\!\left(\sum_{p=1}^{n}\varphi_{qp}(x_p)\right),
 \label{eq:katheorem}
\end{equation}
with univariate inner functions $\varphi_{qp}$ and outer functions $\Phi_q$. Modern Kolmogorov-Arnold networks (KANs) use this result as an organizing principle and replace the fixed scalar edge weights of a conventional multilayer perceptron (MLP) by trainable univariate functions, commonly represented by splines \cite{liu2025kan}. A quick comparison between this two forms of neural networks are shown in table \ref{tab:mlpkan}.

\begin{table}[h]
\caption{Schematic comparison between a conventional dense MLP and the constrained KAN representation used here.}
\label{tab:mlpkan}
\begin{ruledtabular}
\begin{tabular}{lll}
Property & Dense MLP & Present KAN representation\\
\hline
edge object & scalar weight & learned spline function\\
nonlinearity & fixed node activation & trainable edge function\\
interactions & dense layer composition & explicit separable products\\
post-training form & weight/bias tensors & spline coefficients/functions\\
\end{tabular}
\end{ruledtabular}
\end{table}

A dense MLP could therefore be trained for the same regression problem, as demonstrated more generally by neural waveform surrogates but our reason for using the constrained KAN is not an assumption that it must be more accurate. Rather, the learned correction remains in a directly inspectable functional space therefore individual frequency functions, intrinsic-parameter surfaces, and antisymmetric-spin factors can be plotted, differentiated, contribution-ranked, and exported independently of the training framework. This is the specific sense in which the present representation retains more functional information than a generic dense interpolant. It is also complementary to recent physics-informed neural approaches in which known perturbative structure is retained and learned corrections supply missing information \cite{Yoo:2025yhl,Keith:2021}.

\subsection{Physics-factored KAN used in this work}

For either learned residual $F\in\{r,d\}$, with
$d\equiv\delta\ln A$, we use
\begin{align}
	F(x,\lambda)=&\ f_0(u)+g_0(w)
	+\sum_{\rho=1}^{R_u}f_\rho(u)\,
	S_\rho^{(u)}(\widehat\eta,\widehat\chipn)\,
	z_\rho^{(u)}(\widehat\zeta)\nonumber\\
	&+\sum_{\rho=1}^{R_w}g_\rho(w)\,
	S_\rho^{(w)}(\widehat\eta,\widehat\chipn)\,
	z_\rho^{(w)}(\widehat\zeta),
	\label{eq:kanrep}
\end{align}
with $R_u=R_w=6$. Here $\eta$, $\chipn$, and
$\zeta=\delta\chi_a$ are linearly mapped to the normalized coordinates
used by the B-spline bases which is schematically,
\begin{equation}
	\widehat y=
	2\,\frac{y-y_{\min}}{y_{\max}-y_{\min}}-1 .
\end{equation}
The frequency dependence is organized using two coordinates,
\begin{equation}
	u={\cal A}_v\!\left[(\pi x)^{1/3}\right],
	\qquad
	w=\frac{x}{x_{\rm RD}(\lambda)} ,
	\label{eq:clocks}
\end{equation}
where ${\cal A}_v$ is the corresponding affine rescaling of the PN
velocity $v=(\pi Mf)^{1/3}$. We refer to $u$ and $w$ as frequency
\emph{clocks} because they monotonically label progression through the
waveform rather than representing physical time. $u$ organizes the
inspiral in its natural PN variable, whereas $w$ measures frequency
relative to the binary-dependent ringdown scale, with $w\sim1$ near the
fundamental quasinormal-mode frequency. Both clocks enter the same
residual and do not represent separate waveform regions.

Each product term couples a frequency-dependent spline to a
two-dimensional tensor-product spline surface in
$(\widehat\eta,\widehat\chipn)$ and a one-dimensional spline in
$\widehat\zeta$,
\begin{equation}
	\text{frequency function}\times
	\text{intrinsic-parameter surface}\times
	\text{spin-asymmetry function}.
\end{equation}
This structure was arrived at through successive trials of simpler factorizations during model development. In particular, a separable dependence on $\eta$ and $\chipn$ was found to be insufficient, motivating the joint two-dimensional surface, while the additional $\zeta$ factor captures residual two-spin information not represented by $(\eta,\chipn)$ alone. The resulting factorization retains the joint dependence on frequency, mass ratio, and spin while keeping the learned representation explicit and low dimensional.

\begin{figure}[t]
\centering
\includegraphics[width=0.96\linewidth]{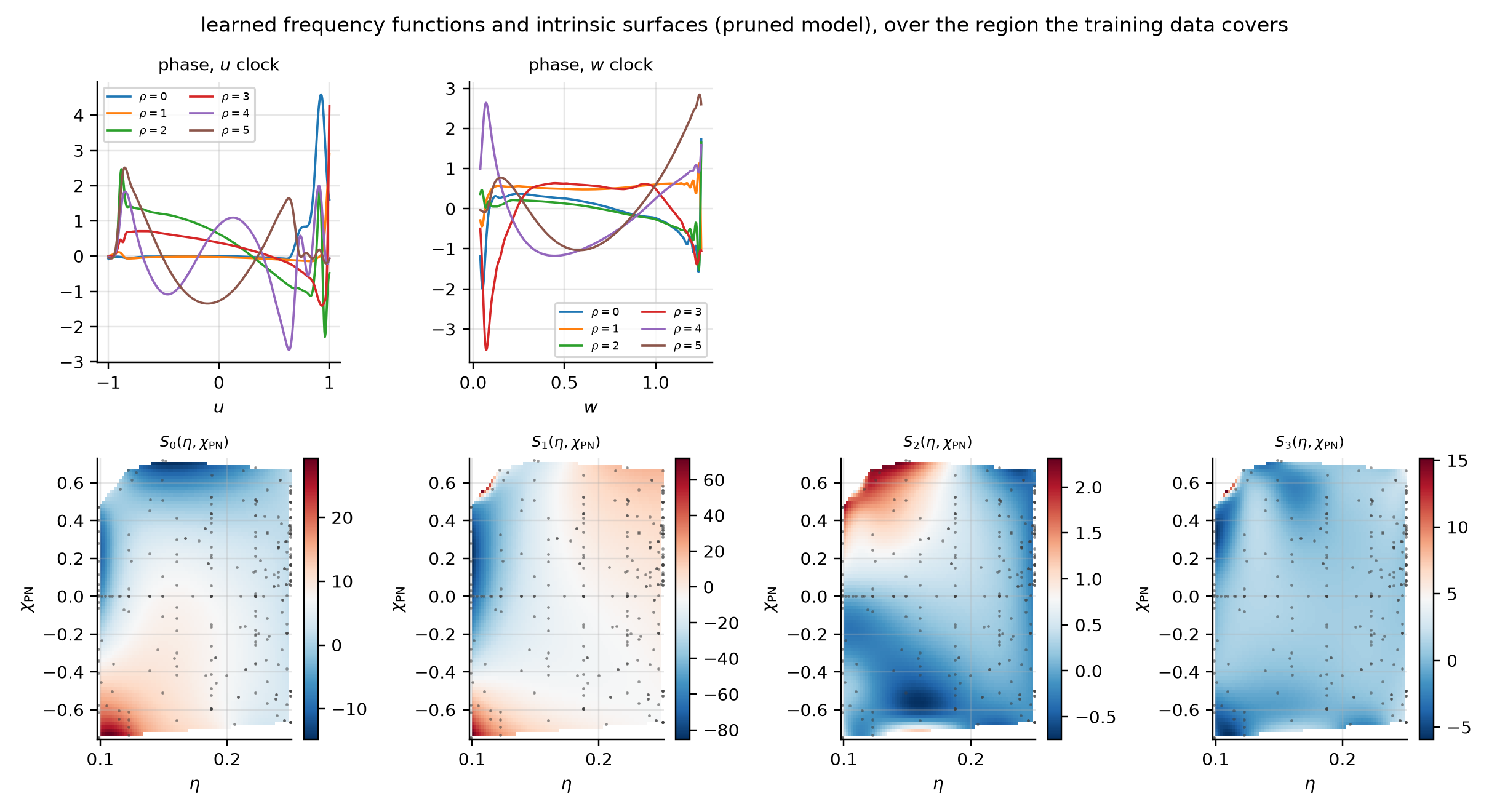}
\caption{Examples of learned factors in the final rank-6 KAN representation.  Top: phase-frequency functions associated with the PN-scaled $u$ and ringdown-scaled $w$ clocks.  Bottom: representative intrinsic surfaces $S_\rho(\eta,\chipn)$ with training configurations overlaid.  The figure is intended to show the explicit functional organization of the learned correction, not to assign an independent physical interpretation to any single factor.}
\label{fig:functions}
\end{figure}

All one-dimensional functions are represented by open, clamped cubic
B splines, while each $S_\rho$ is a tensor-product cubic B-spline
surface. Repeated knots at the boundaries keep the basis well defined
over the full spline interval and ensure non-negativity, exact
representation of constants, and partition of unity up to the
endpoints. The production or final representation uses
\begin{equation}
 n_u=96,
 \qquad n_w=80,
 \qquad n_\theta=8,
 \qquad R_u=R_w=6.
\end{equation}
For one residual branch the additive terms contribute $96+80$ coefficients, the $u$ products contribute $6(96+8^2+8)$, and the $w$ products contribute $6(80+8^2+8)$.  With independent phase and logarithmic-amplitude branches, the complete model therefore contains
\begin{equation}
 2\left[96+80+6(96+64+8)+6(80+64+8)\right]
 =\boxed{4192}
 \label{eq:paramcount}
\end{equation}
trainable coefficients and 24 complete spline-product terms. Figure~\ref{fig:functions} illustrates that after training, the frequency factors and intrinsic surfaces remain explicit functions rather than hidden combinations of dense network weights as was the motivation for using KAN.

The explicit product structure also gives a natural unit for model reduction. Instead of pruning individual scalar coefficients, we test whether an entire contribution of the form $f_\rho S_\rho z_\rho$ or $g_\rho S_\rho z_\rho$ can be removed without degrading validation accuracy. The pruning stage therefore asks whether the learned correction can be represented with fewer complete functional terms, while the held-out test set remains untouched.

\section{Training objective and optimization}
\label{sec:training}

\subsection{Masked, gauge-invariant objective}

Training is performed in full batch over the 280 training systems. Each numerical waveform contributes only within its trustworthy frequency interval, implemented through the mask $m_i(x)$. On the logarithmic frequency grid we combine three complementary weighting channels. To suppress poorly constrained very-low-frequency information without introducing a sharp boundary, we use a smooth $C^2$ roll-on
\begin{equation}
	R(x)=s^3(10-15s+6s^2),
	\qquad
	s=\mathrm{clip}\!\left(
	\frac{x-0.0059}{0.0070-0.0059},0,1\right).
	\label{eq:rollon}
\end{equation}
The lower boundary $x=0.0059$ corresponds to $20\,{\rm Hz}$ for a $60M_\odot$ binary, while the short transition to $x=0.0070$ avoids a discontinuous optimization weight. We then define
\begin{align}
	W_i^{\rm snr}(x)&\propto
	a_{{\rm NR},i}^2(x)\,x\,R(x)m_i(x),\\
	W_i^{\rm mrg}(x)&=
	W_i^{\rm snr}(x)
	\Theta\!\left(\frac{x}{x_{{\rm RD},i}}-0.5\right),\\
	W_i^{\rm uni}(x)&\propto
	m_i(x)\,\Theta[R(x)>0].
	\label{eq:weights}
\end{align}
Each channel is normalized to mean unity over the retained samples of each simulation. The first follows the flat-noise overlap density on a logarithmic grid, for which $df=f\,d\ln f$ gives a weight proportional to $a_{\rm NR}^2x$. The merger channel applies the same weighting only for $x/x_{\rm RD}\geq0.5$, preventing the shorter strong-field region from being overwhelmed by the inspiral. The uniform channel gives comparable importance
to all retained frequencies and prevents the fit from being determined only by the largest-amplitude samples.

For the phase residual, arbitrary relative time and phase shifts must not
contribute to the loss. For $e_{r,i}=r_{\rm KAN}-r_{\rm NR}$, we therefore subtract, separately for each system and weighting channel, its weighted least-squares projection onto $\{1,2\pi x\}$. Denoting the remaining error by $Pe_r$, the data terms are
\begin{align}
	\mathcal L_{\phi}&=
	2\langle W^{\rm snr}(Pe_r)^2\rangle+
	2\langle W^{\rm mrg}(Pe_r)^2\rangle+
	\langle W^{\rm uni}(Pe_r)^2\rangle,\\
	\mathcal L_A&=
	\langle W^{\rm snr}e_A^2\rangle+
	\langle W^{\rm mrg}e_A^2\rangle+
	\langle W^{\rm uni}e_A^2\rangle,
	\qquad
	e_A=d_{\rm KAN}-d_{\rm NR}.
	\label{eq:lossdata}
\end{align}
The complete training objective is
\begin{equation}
	\mathcal L=
	\mathcal L_\phi+\mathcal L_A
	+10^{-5}\mathcal R_{1{\rm D}}
	+0.3\mathcal R_{\rm surf}
	+10^{-4}\mathcal R_{\rm sparse}.
	\label{eq:loss}
\end{equation}
The regularizers act directly on the spline coefficients. For a one-dimensional spline with coefficients $c_k$, we define $\Delta^2c_k=c_{k+2}-2c_{k+1}+c_k$ and use
\begin{equation}
	\mathcal R_{1{\rm D}}
	=
	\sum_{q\in{\cal S}_{1{\rm D}}}
	\left\langle
	\bigl(\Delta^2c^{(q)}\bigr)^2
	\right\rangle ,
	\label{eq:r1d}
\end{equation}
where ${\cal S}_{1{\rm D}}$ contains the additive frequency splines and the frequency- and $\zeta$-dependent splines entering the product terms. For each intrinsic surface with coefficients $C^\rho_{jk}$,
\begin{equation}
	\mathcal R_{\rm surf}
	=
	\sum_{\rm blocks,\rho}
	\left[
	\left\langle(\Delta_j^2C^\rho_{jk})^2\right\rangle+
	\left\langle(\Delta_k^2C^\rho_{jk})^2\right\rangle
	\right],
	\label{eq:rsurf}
\end{equation}
which suppresses unnecessary variation between sparsely sampled intrinsic configurations. Finally,
\begin{equation}
	\mathcal R_{\rm sparse}
	=
	\sum_{\rm blocks,\rho}
	\left[
	\frac{1}{n_\rho}\sum_k
	\bigl(c_k^{(\rho)}\bigr)^2+10^{-12}
	\right]^{1/2},
	\label{eq:rsparse}
\end{equation}
where $c_k^{(\rho)}$ are the coefficients of the frequency spline associated with each rank term. This provides a weak structured sparsity bias, while the subsequent model-reduction stage tests removal of complete product terms directly. No additional low-frequency output anchor is used in the production model. Low-frequency control instead comes from the analytic backbone, the per-system validity masks, and the smooth optimization roll-on.

\subsection{Warm start}

Equation~\eqref{eq:kanrep} is nonlinear in products of spline factors, and a fully random initialization would require the optimizer to discover simultaneously the residual shape, its parameter dependence, and the relative scales of the factors. We therefore construct a deterministic mask-aware rank-$4$ warm start, where the rank denotes the number of
frequency-intrinsic-parameter product terms in each of the $u$ and $w$ branches. Weighted linear least squares first determines the additive $f_0(u)+g_0(w)$ contribution. The remaining masked residual is then used to initialize four dominant product terms in the $u$ branch by a low-rank factorization, followed by the corresponding construction on a common $w$ grid. Frequencies outside each SXS waveform's trustworthy support do not enter these factorizations.

For a higher-rank model, the rank-$4$ solution is embedded exactly and the additional product terms are initialized with zero contribution while retaining nonzero gradients. Thus the final rank-$6$ model begins from the same deterministic waveform rather than from a rank-dependent random initialization. The warm start only initializes the optimization, and all $4192$ production coefficients remain trainable.

\subsection{Optimizer, checkpoint selection, and seed stability}

The production rank-$6$ model is trained with Adam for $4000$ epochs. The frequency-spline coefficients use an initial learning rate $3\times10^{-2}$, while all remaining parameters start at $3\times10^{-3}$. A cosine schedule decreases both toward a common floor of $6\times10^{-5}$. Validation diagnostics are evaluated every $25$ epochs, and within each run we retain the checkpoint with the smallest median validation mismatch proxy. Using the median rather than the mean reduces the sensitivity
of model selection to a small number of difficult validation configurations.
\begin{table}[t]
	\caption{Production training configuration.}
	\label{tab:train}
	\begin{ruledtabular}
		\begin{tabular}{ll@{\qquad}ll}
			Quantity & Value & Quantity & Value\\
			\hline
			training/validation systems & $280/75$ & epochs & $4000$\\
			$R_u=R_w$ & 6 & $n_u,n_w,n_\theta$ & $96,80,8$\\
			trainable coefficients & 4192 & optimizer & Adam\\
			initial base learning rate & $3\times10^{-3}$
			& initial frequency-spline rate & $3\times10^{-2}$\\
			cosine learning-rate floor & $6\times10^{-5}$
			& validation cadence & 25 epochs\\
			phase channel weights & $(2,2,1)$
			& amplitude channel weights & $(1,1,1)$\\
			$\mathcal R_{1\rm D}$ weight & $10^{-5}$
			& $\mathcal R_{\rm surf}$ weight & $0.3$\\
			$\mathcal R_{\rm sparse}$ weight & $10^{-4}$
			& low-frequency anchor weight & $0$\\
			training seeds & $0,1,2,3,4$ & production seed & $2$\\
		\end{tabular}
	\end{ruledtabular}
\end{table}

To quantify initialization dependence, the complete rank-$6$ training is repeated for five seeds fixed before model construction. Their true time- and phase-maximized validation median mismatches are $3.76,3.93,3.83,4.33,$ and $3.50\times10^{-5}$ for seeds $0-4$. Rather than selecting the numerically best realization, we choose seed $2$, whose validation median is the median of the five runs, as the production checkpoint. As a compactness check, the same five-seed study at rank $7$ gives a median-across-seeds mismatch of $3.33\times10^{-5}$, compared with $3.83\times10^{-5}$ at rank 6, while increasing the number of trainable coefficients from $4192$ to $4832$. We therefore retain the predefined rank-$6$ representation.

\begin{figure}[t]
\centering
\includegraphics[width=0.88\linewidth]{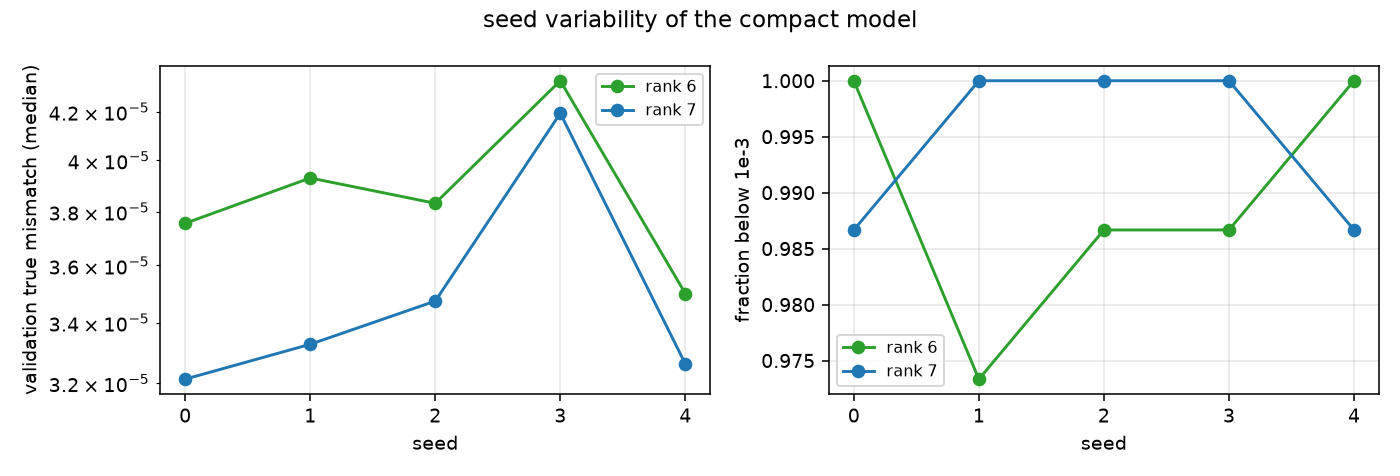}
\caption{Seed stability used as an optimization diagnostic.  The production checkpoint is not the numerically best seed; seed 2 is chosen because its validation median is the median of the five rank-6 runs.  Rank 7 is shown only as a capacity check.}
\label{fig:seeds}
\end{figure}

All training reported here was performed on an Apple M1 MacBook using the PyTorch Metal Performance Shaders (MPS) backend.  The execution log records approximately 37 minutes for the complete five-seed rank-6 study (about 7--8 minutes per 4000-epoch run).  Including the five rank-7 capacity-check runs brings the cumulative training study to approximately 78 minutes.

\section{From the trained KAN to a deployable waveform representation}
\label{sec:deploy}

The production rank-$6$ checkpoint contains $24$ complete spline-product
terms. We test whether this representation can be simplified without
materially degrading its validation accuracy. The terms are ordered by their
mask-weighted RMS contribution, with the phase and amplitude branches treated
independently. After removing a candidate term, the remaining model is refitted
for $800$ epochs from the trained state. A deletion is accepted only if both
the median and the $90$th-percentile validation mismatch remain within $15$\%
of their values for the full representation.

No candidate deletion satisfies this criterion. The first attempted amplitude
deletion increases the validation median from $3.83\times10^{-5}$ to
$1.06\times10^{-4}$ and the $90$th percentile from $1.13\times10^{-4}$ to
$4.73\times10^{-4}$. The first attempted phase deletion gives
$1.31\times10^{-4}$ and $4.85\times10^{-4}$ for the same two quantities.
The final waveform model therefore retains all $24$ spline-product terms.

\begin{figure}[t]
\centering
\includegraphics[width=0.92\linewidth]{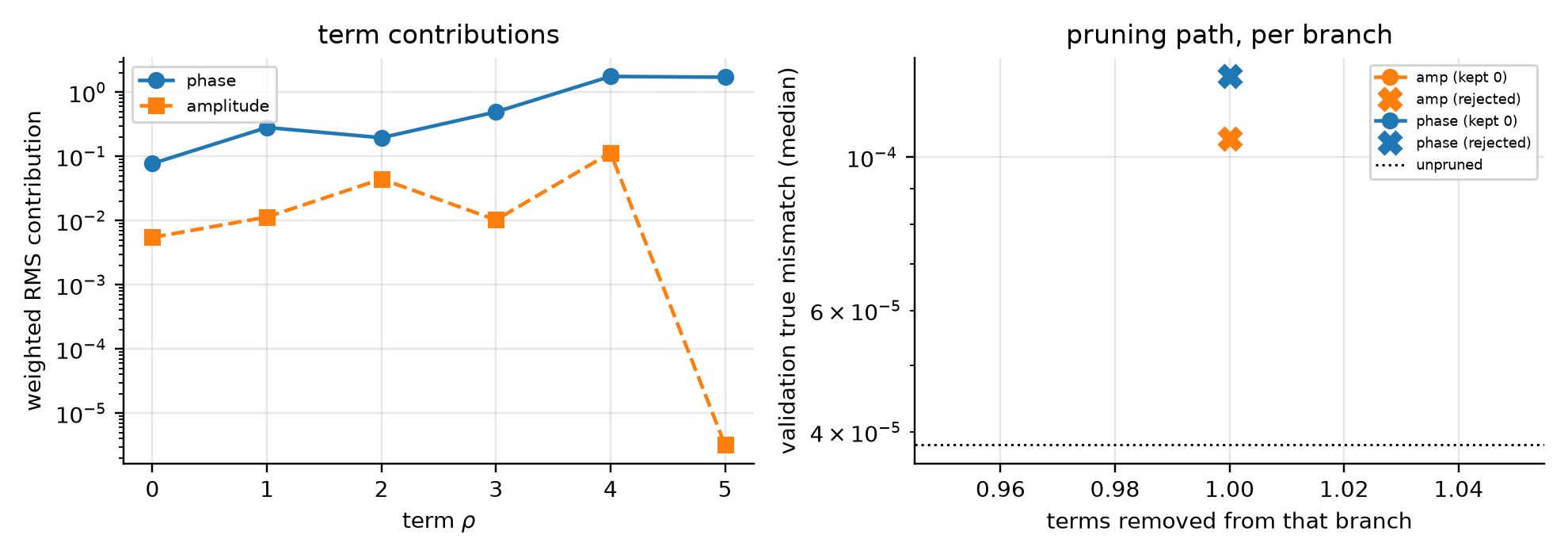}
\caption{Validation-controlled model-reduction diagnostic.  Complete product terms are ranked by their contribution, and the first trial deletion in each residual branch is rejected because the deletion-plus-refit result violates the 15\% acceptance tolerance in the validation median and/or 90th percentile. The final KAN therefore retains all 24 product terms.}
\label{fig:pruning}
\end{figure}

This should be interpreted as a result of the model-reduction procedure used
here rather than as a statement of mathematical irreducibility. Within the
adopted deletion, refitting, and validation criterion, none of the complete
product terms can be removed while preserving the accuracy of the trained
representation. Figure~\ref{fig:pruning} records this model-reduction test. 
\begin{figure}[t]
	\centering
	\includegraphics[width=0.92\linewidth]{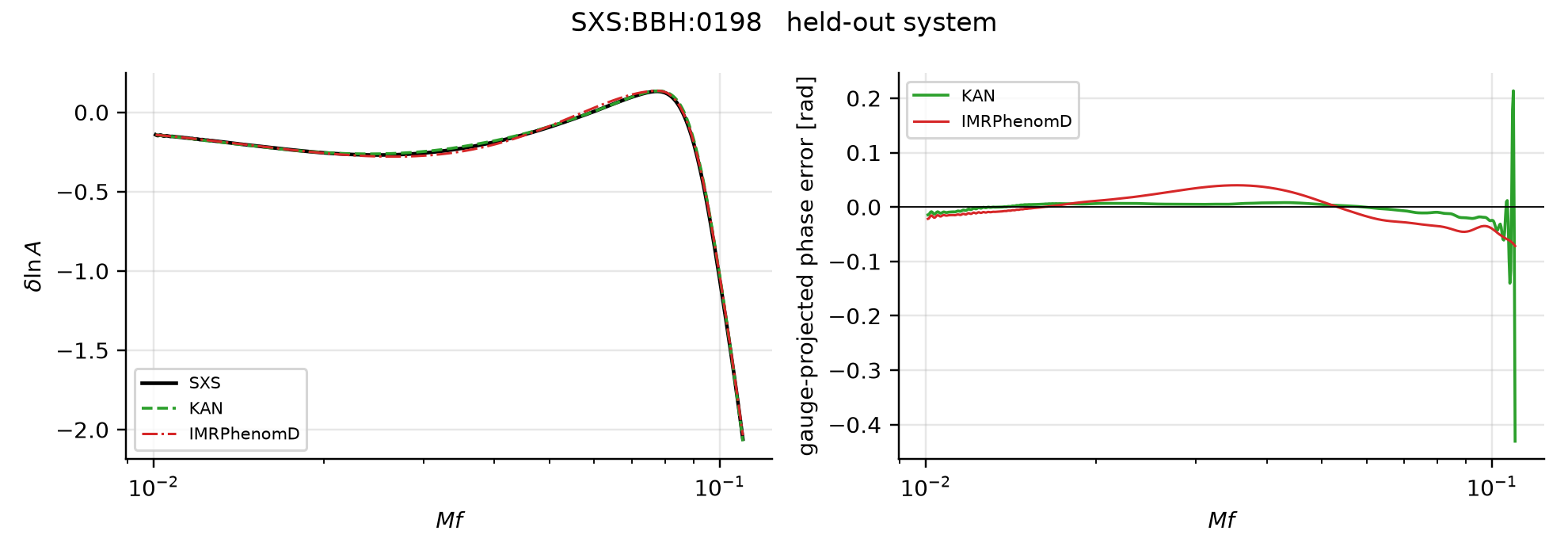}
	\caption{Representative held-out waveform comparison. The left panel shows
		the logarithmic-amplitude correction relative to the Newtonian amplitude for
		SXS, the KAN model, and IMRPhenomD. The right panel shows the corresponding
		gauge-projected phase errors relative to SXS.}
	\label{fig:residual}
\end{figure}
The held-out $75$-system interpolation test is evaluated only after the representation is fixed. The final KAN gives a median flat-noise mismatch $2.73\times10^{-5}$, a $90$th percentile $6.51\times10^{-5}$, and a maximum $2.46\times10^{-4}$; all 75 held-out systems lie below $10^{-3}$.  Figure~\ref{fig:residual} shows one held-out waveform in the residual variables. 
Finally, the trained spline coefficients are exported to a pure-NumPy evaluator so waveform generation no longer depends on PyTorch or the training code.  At $12,800$ random in-domain points, the double-precision NumPy and Torch evaluations agree to $6.55\times10^{-10}$ rad in the phase residual and $3.95\times10^{-11}$ in $\dlnA$.  A separate frequency-time-frequency consistency test gives relative error $6.27\times10^{-16}$ over the interior band. The deployable code and coefficients are provided in the project repository \cite{ChattopadhyayKasthaKAN2026}.



\end{document}